\documentclass[journal,transmag]{IEEEtran}

\pdfoutput=1

\ifCLASSINFOpdf
\else
\fi

\usepackage{url}
\usepackage[colorlinks=true,urlcolor=blue,linkcolor=blue,citecolor=red]{hyperref}
\newcommand{\URL}[1]{{{\url{#1}}}}

\newcommand{\vect}[1]{{\mbox{\boldmath{$\bf #1$}}}}
\newcommand{\Vx}{{\vect{x}}}
\newcommand{\Vy}{{\vect{y}}}

\newcommand{\Vv}{{\vect{v}}}
\newcommand{\Vk}{{\vect{k}}}

\newcommand{\norm}[1]{{\left|\left|{#1}\right|\right|}}
\newcommand{\sizeOfSet}[1]{{\norm{x}}}
\newcommand{\adjoint}[1]{{{#1}^{\top}}}
\newcommand{\distance}[2]{{\mbox{dist}({#1},{#2})}}

\newcommand{\logHEC}{{\log(\,\mbox{HEC}\,)}}
\newcommand{\logPHI}{{\log(\,\mbox{PHI}\,)}}
\newcommand{\logPCI}{{\log(\,\mbox{PCI}\,)}}

\usepackage{graphicx}

\usepackage{xcolor}

\begin{document}

\title{Inferring Microclimate Zones from Energy Consumption Data}

\author{\IEEEauthorblockN{Thuy Vu\IEEEauthorrefmark{1}, and D.S. Parker\IEEEauthorrefmark{1}}
\IEEEauthorblockA{\IEEEauthorrefmark{1}Computer Science Department,
University of California, Los Angeles, CA 90095 USA}
\thanks{Manuscript received September 7, 2018.  
Corresponding author: D.S. Parker (email: stott\@cs.ucla.edu).
}}


\markboth{Microclimate Zones from Energy Consumption Data}%
{Microclimate Zones from Energy Consumption Data}

\IEEEtitleabstractindextext{%
\begin{abstract}

Climate zones are an established part of urban energy management.
California is divided into 16 climate zones, for example,
and each zone imposes different energy-related building standards ---
so for example different zones have different roofing standards.
Although developed long ago, these zones continue to shape urban policy.
New climate zone definitions are now emerging in urban settings.
Both the County of Los Angeles and the US Department of Energy have recently adopted refined zones ---
for restructuring electricity rates, and for scoring home energy efficiency.
Defining new zones is difficult, however, because climates depend on variables that are difficult to map \cite{UncertaintyOfMicroclimateVariables}.

In this paper we show that Los Angeles climate zones can be inferred from energy use data.
We have studied residential electricity consumption (EC) patterns in Los Angeles data.
This data permits identification of geographical zones
whose EC patterns are characteristically different from surrounding regions.
These regions have topographic boundaries in Los Angeles consistent with microclimates.

Specifically, our key finding is that
\emph{EC-microclimate zones}
--- regions in which block groups have similar Electricity Consumption patterns over time ---
resemble environmental microclimate zones.
Because they conform to microclimates, but are based directly on EC data,
these zones can be useful in urban energy management.

We also show how microclimates and household electricity consumption in Los Angeles are strongly linked to
socioeconomic variables like income and population density.
These links permit data-driven development of microclimate zones that support zone-specific modeling and improved energy policy.

\end{abstract}

}

\maketitle

\IEEEdisplaynontitleabstractindextext

\IEEEpeerreviewmaketitle

\section{Introduction}

\IEEEPARstart{U}{rban}
energy management is an important problem facing all cities.
In this paper,
we describe some results of analyses on residential electricity consumption (EC),
using monthly consumer data from
regional energy utilities, including the Los Angeles Department of Water and Electricity (LADWP),
SoCal Edison, SoCal Gas, and others.
Specifically, we focus here on analysis of 
block group totals, giving a summary of EC in small Census-defined neighborhoods.

This increasing availability of data makes it possible to go beyond traditional accounting,
and to perform more comprehensive analysis of patterns and processes in energy systems.
The Los Angeles electricity consumption data offers examples of the importance of
resource flows \cite{ExpandedUrbanMetabolism}
and of climate \cite{UrbanClimatology}
in urban design,
showing how better understanding of both can prove useful in policy.
In particular, we show that it gives insight into connections between EC and
microclimates \cite{Microclimates}.

\subsection{Climate Zones and Energy Policy}

Energy use depends on climate.
As a result, in areas like California with diverse climates,
climate zones have important roles in energy management.
For example, the California Energy Commission has divided the state into
16 Energy Climate Zones \cite{ClimateZones,MoreOnClimateZones}, such as
for coastal Los Angeles (\#6), Long Beach/El Toro (\#8), and inland Los Angeles/Pasadena (\#9).
Different energy-related building standards are imposed in each zone, such as for HVAC or roofing.

\begin{figure}[htbp]
\centering
\vspace{-0.1in}
\includegraphics[width=3.1in]{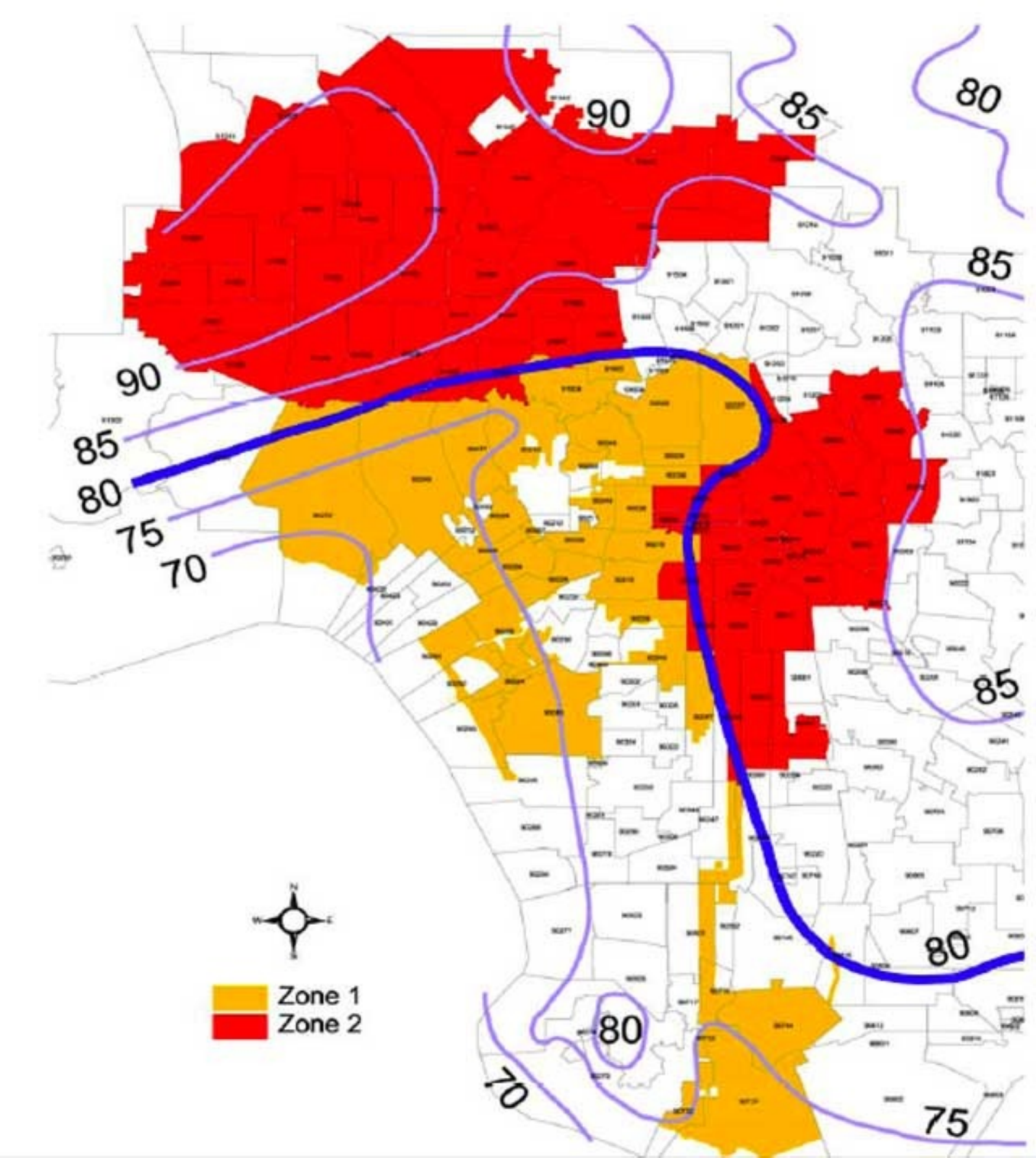}

\caption{
About five years ago,
two climate zones were proposed by LADWP for restructuring its rates
\cite{LADWPRateRestructuring}, so that consumers in the warmer (and less affluent)
Zone 2 paid relatively lower rates.
Also, three tiers of use were identified in each zone, with consumers using more kWH paying higher rates.
Although they reflect `microclimates' (like the zones in this paper),
isotherm boundaries do not define sharp zones, and this proposal was controversial.
Zone 1 is somewhat similar to California's Energy Climate Zone \#6, but not equivalent.
\label{UCLA_microclimate_map}
}
\end{figure}

Climate zones also have been used in defining rates, which are vital controls on electricity consumption.
LADWP, in its rate restructuring about ten years ago \cite{LADWPRateRestructuring},
defined two microclimate zones based on historical temperature averages.
These zones are shown in Figure~\ref{UCLA_microclimate_map};
Zone 1 (to the west) has higher electricity rates.

Although these zones were defined with energy use in mind,
they will need to evolve to remain in step with rapid growth in these areas.
Finer zone granularity
in Los Angeles could aid in improving energy management.

\begin{figure}[htbp]
\centering
\hspace{-0.01in}\includegraphics[width=3.3in]{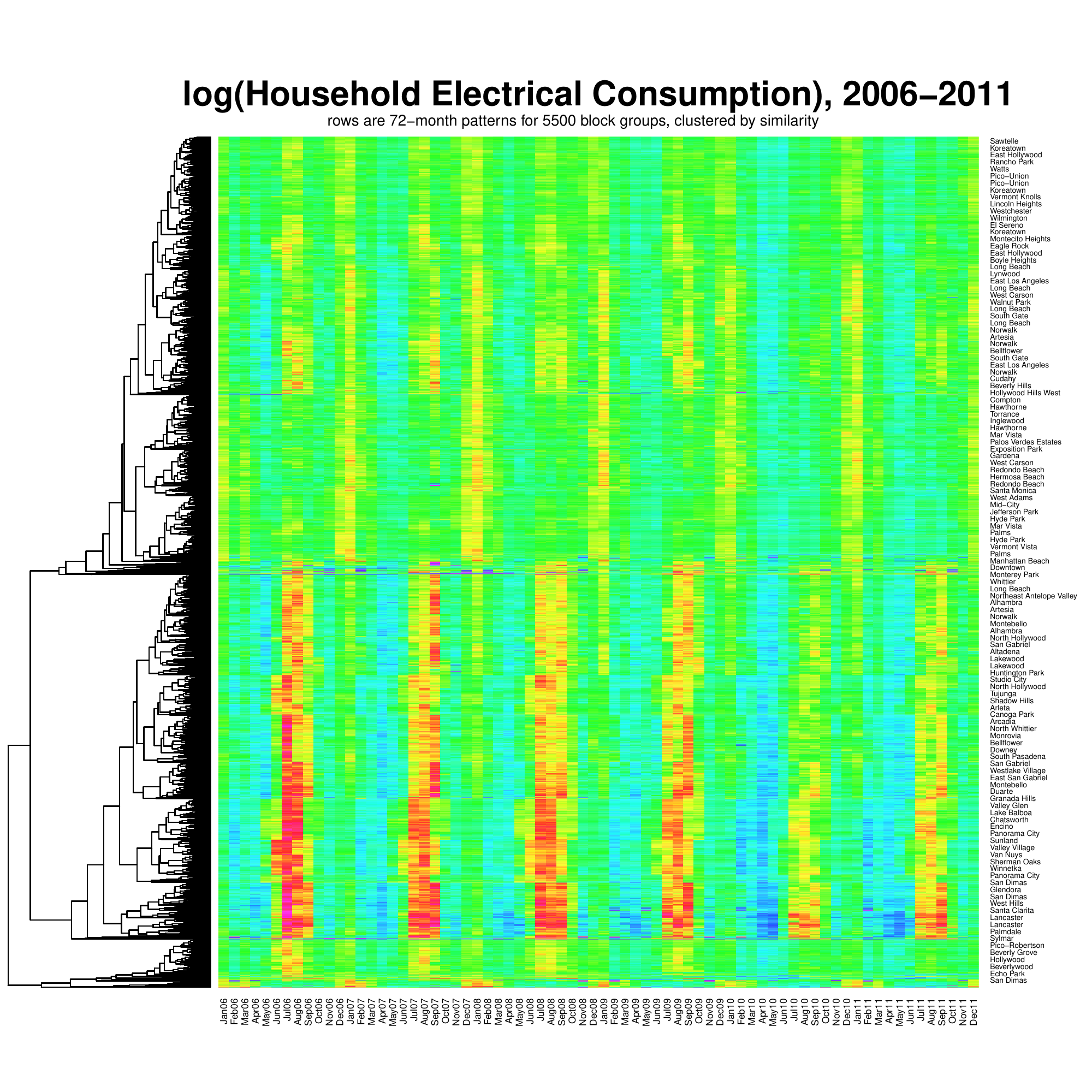}

\vspace{-0.2in}
\caption{
A visual display of aggregate Household Electricity Consumption (HEC) data used in this study, showing
the history of residential EC in Los Angeles County over the six years 2006-2011.
Specifically, this heatmap shows $\logHEC$ at the block group level
for the 72 months in this period, with each row normalized to sum to 1.
The display is thus a 5000$\times$72 summary of electricity consumption,
with rows representing block groups.
Clear seasonal patterns are visible, and hierarchical clusters of neighborhoods with similar patterns are evident.
Neighborhood names are displayed on the right for about every 50th row, showing that
climate and geography are both important factors in Los Angeles electricity consumption.
However, only $\logHEC$ values were used to produce this graphic
--- no geographic or climate data was used at all.
At least 3 different patterns/clusters are apparent, suggesting that
there are distinct EC patterns in at least 3 geographic zones.
\label{logHEC_heatmap}
}
\end{figure}

Faced with increasing complexity of consumer demand and energy supply options,
the California Energy Commission and state-wide electricity providers
evolved the California IEPR (Integrated Energy Policy Report) over the past decade
(e.g., \cite{IEPR2013,IEPR2014,IEPR2015,IEPR2016,IEPR2017,IEPR2018}).
It has been `a biennial integrated energy policy report that assesses major energy trends and issues facing the state's electricity, natural gas, and transportation fuel sectors and provides policy recommendations to conserve resources; protect the environment; ensure reliable, secure, and diverse energy supplies; enhance the state's economy; and protect public health and safety'.
In IEPR workshops pressing electricity issues are reviewed, with significant impact on urban policy and action plans.
Many topics focus on climate.

\medskip

Recently the state of California embarked on an ambitious plan to improve efficiency:
Assembly Bill 758 \cite{AB758} 
requires the California Energy Commission,
California Public Utilities Commission,
and other stakeholders
to develop a comprehensive program to increase energy efficiency in existing buildings.
An action plan was passed in August 2015 \cite{AB758_Plan_2015}
to double energy efficiency savings in retail gas and electricity use
by 2030.
The December 2016 update \cite{AB758_Plan_Update_2016} reports on
complex econometric forecasting models, and
strategies that can contribute to this requirement.
Progress is being tracked dynamically \cite{TrackingProgress}.
New kinds of energy policy are now needed.

Very recently, this ambitious plan became more ambitious.
In May 2018, California required solar energy in \emph{all} new residential construction starting in 2020
\cite{CaliforniaMandatesSolar,CaliforniaMandatesSolar3}.
The 2020 energy code also will require greater insulation, window, and appliance efficiencies.
In August 2018, California also approved a bill requiring California to generate \emph{100\%} of its energy from carbon-free sources by 2045
\cite{CarbonFreeEnergy}.
Furthermore it passed a mandate to generate at least \emph{60\%} of its energy from renewable resources by 2030.
These plans require rapid innovations in energy management.

\subsection{The Key Idea: Data-driven Microclimate Zones}

In \cite{AAAI_Microclimates} we showed that
residential electricity consumption data can be sufficient to {infer climate}.
Briefly: \emph{home electricity consumption data includes a history of climate.}

As a result, we were able to identify \emph{microclimates} --- regional climate patterns --- from residential Electricity Consumption (EC) data
for Los Angeles.
Although these `EC-microclimate zones' were obtained with clustering,
they resemble environmental microclimate zones.

Furthermore, a data-driven approach permits the incorporation of quantitative models into policy.
In our data, EC-microclimates are strongly linked with economic variables,
highlighting connections between climate, economics, and energy use.
Existing models can be sharpened for microclimates,
and support data-driven energy policy.

\subsection{Overall Findings}
\label{Results}

Climate zones have well-established roles in energy policy \cite{ClimateZones},
and Los Angeles is interesting 
in its diversity of microclimates.
We were interested in determining how
microclimates are related to residential Electricity Consumption (EC).

Some central findings for our data were as follows:

\begin{itemize}
 \item {EC data can be used to identify zones with similar EC patterns}.
 \item {These zones appear strongly related to climate}, so we call them \emph{EC-microclimate zones}.
 \item {These EC-microclimate zones can be validated as corresponding to climate}.
 \item {EC-microclimate zones are also strongly linked to income},
suggesting that they could be useful in energy-related models and policy.
 \item more accurate large-scale models for EC can be developed by
{developing different models for each EC-microclimate zone}.
\end{itemize}
{Findings for household electricity consumption (HEC):}
\begin{itemize}
 \item HEC patterns
are highly region-specific geographically, in ways related to climate.
 \item in each EC-microclimate zone, $\logHEC$ follows a GPRF (Gaussian Process Random Field).
 \item $\logHEC$ is highly correlated with socioeconomic variables like income. 
 \item Energy demand ($\logHEC$) forecasts can be improved by building separate models in each microclimate zone.
\end{itemize}

\begin{figure}[htbp]
\centering
\vspace{-0.25in}
\includegraphics[width=3.20in]{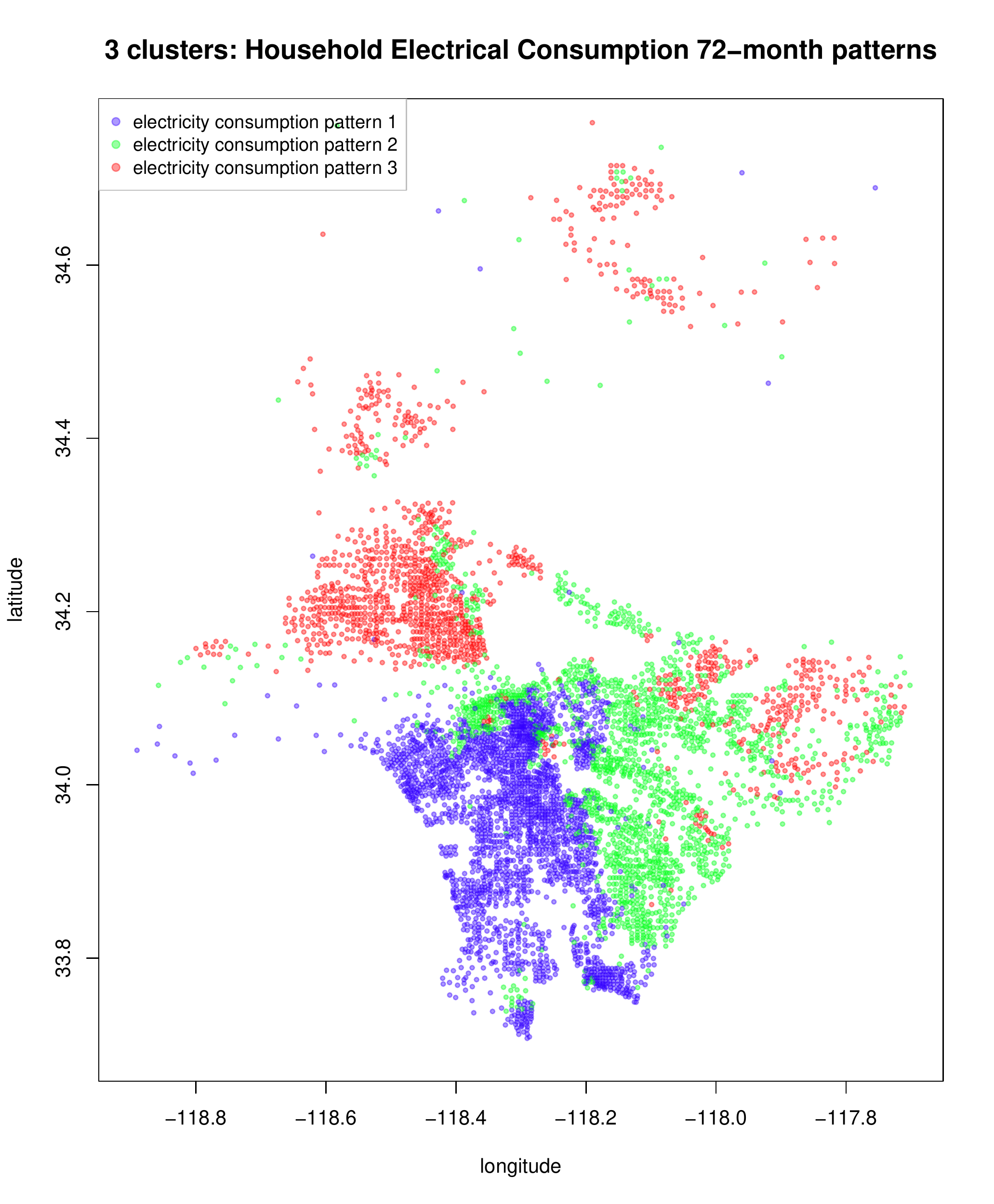}
\caption{A map of Los Angeles, showing about 5000 individual block groups as colored dots
positioned at the geographic center of the block group.
The western and southern coastlines are visible as the outline of the blue region at the bottom.
The three zones indicated by colors correspond to three different electricity consumption (EC) patterns.
These zones --- the western Los Angeles basin, the northern San Fernando Valley, and the eastern Los Angeles area / San Gabriel Valley ---
also appear to correspond to different microclimates.
The direct connection between these zones and Los Angeles County topography is visible in
Figure~\ref{HouseholdIncomeMap}.
The zones were not determined by geography;
they reflect only the similarity of EC patterns over 2006-2011.
As they appear to reflect climate, we refer to them as \emph{EC-microclimate zones}.
\label{basic_electricity_consumption_map}
}
\end{figure}

\begin{figure}[htbp]
\centering
\vspace{-0.15in}
\includegraphics[width=2.8in]{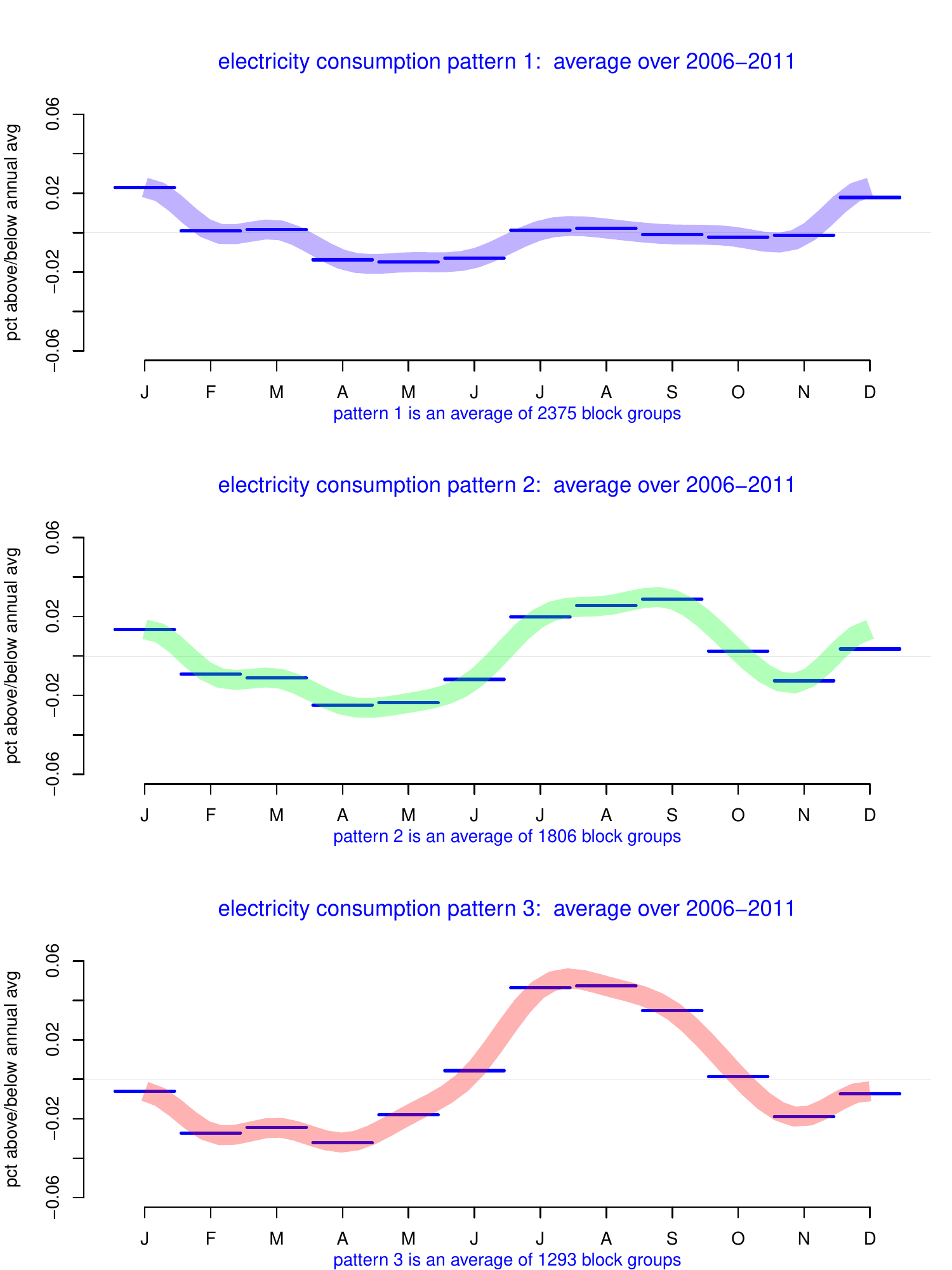}
\caption{
Annual household electricity consumption patterns are shown
for the three zones shown in Figure~\ref{basic_electricity_consumption_map}.
In our analysis, a \emph{pattern} is a sequence of 72 monthly values (over the six years),
and these plots show the average 12-month cycle.
Zones 2 and 3 exhibit higher electricity consumption in Summer,
while zone 1 exhibits highest consumption in Winter.
Zones 2 and 3, representing eastern and northern Los Angeles County (and the San Gabriel and San Fernando Valleys),
also exhibit higher annual variability, corresponding to inland locations.
Zone 1 has much lower variability, corresponding to the western Los Angeles basin and coastal areas.
This plot shows the primary three clusters (indicated by color) of similar annual EC patterns,
plotted at the location of each block group.
The clustering made {no} use of location information;
it considered only the $\logHEC$ time series over the 6 years (normalized to sum to 1),
where \textit{HEC = EC / households}, electricity consumption per household.
In other words, only the similarity of 72-month $\logHEC$ patterns
determined EC-microclimate zone structure.
\label{basic_electricity_consumption_plots}
}
\end{figure}

\begin{figure*}[htbp]
\centering
\vspace{-0.25in}
\begin{minipage}{7.0in}
\includegraphics[width=3.4in]{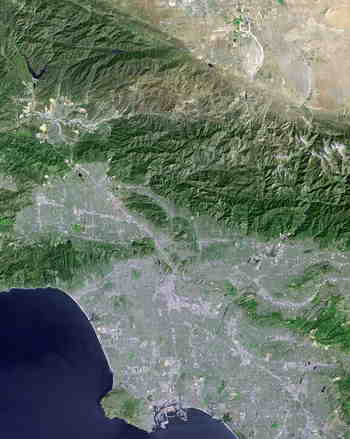}
\includegraphics[width=3.6in]{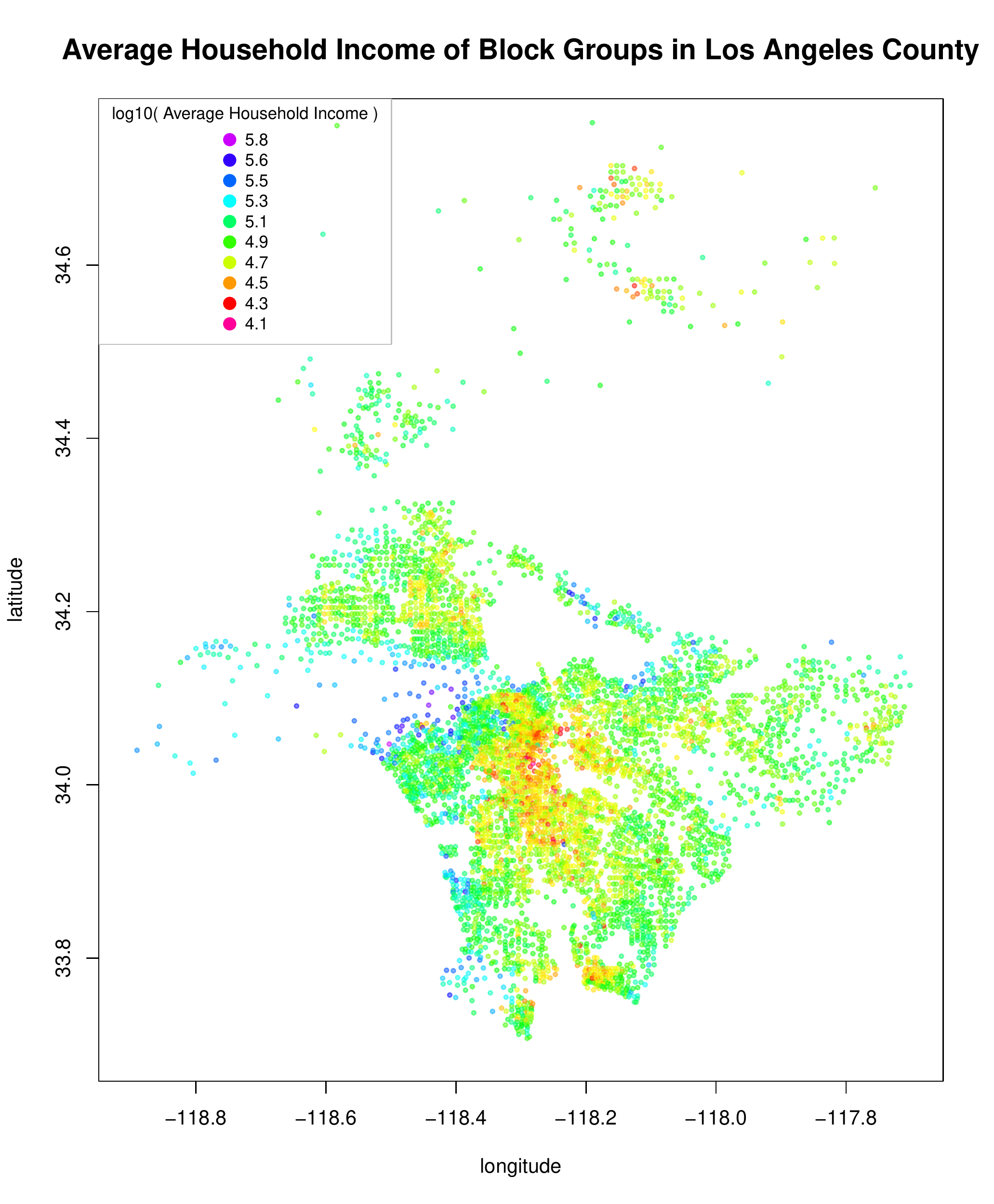}
\end{minipage}

\caption{
The satellite image on the left shows Los Angeles topography.
The map on the right shows
Household Income (Per Household Income, PHI) of block groups in Los Angeles County.
PHI is highest in coastal and hillside areas.
These areas have desirable microclimates, with lower temperatures, higher wind, higher relative humidity,
and more vegetation.
They are also characterized by high electricity consumption and high water use \cite{WaterConsumptionPolicy}.
\label{HouseholdIncomeMap}
}
\end{figure*}

\bigskip

The most important
finding was that clustering of time series of residential electricity consumption (EC)
identified geographic regions of similar patterns that appear related to climate.
These \emph{EC-microclimate zones}, inferred from EC data, have many potential applications in energy managment.

\section{Electricity Consumption Data Analysis}

We analyzed data from the
California Center for Sustainable Communities (CCSC)
containing 2006-2011 monthly electricity consumption histories for Los Angeles County.
An earlier analysis of electricity consumption per capita in California \cite{PerCapitaUse}
showed both the promise and difficulty of accurate modeling, and the importance of model choice in obtaining forecasts.
To gain an understanding of electricity consumption,
we analyzed data from about 5000 block groups
(census-related street-level geographic regions).

In this paper, for insight into residential electricity consumption patterns,
we consider aggregative results for residential block groups in Los Angeles County.
Electricity consumption in a block group depends on the number of consumers (`households'),
and the ratio measures {household electricity consumption}.
Our analysis here
is based on monthly consumption and household count values in individual block groups.
The data ranges can vary by orders of magnitude.
We study logarithmically-scaled consumption,
which turns out to be approximately normally distributed:
\begin{eqnarray*}
\logHEC & = & \log(\,\mbox{Household Electricity Consumption}\,) \\
 & = & \log(\,\mbox{Electricity Consumption} ~/~ \mbox{Households}\,) .
\end{eqnarray*}
Here $\log = \log_{10}$; all logarithms are to base 10 in this paper.

\subsection{Overview of the Data}

\begin{figure*}[htbp]
\centering
\vspace{-0.25in}
\includegraphics[width=3.05in]{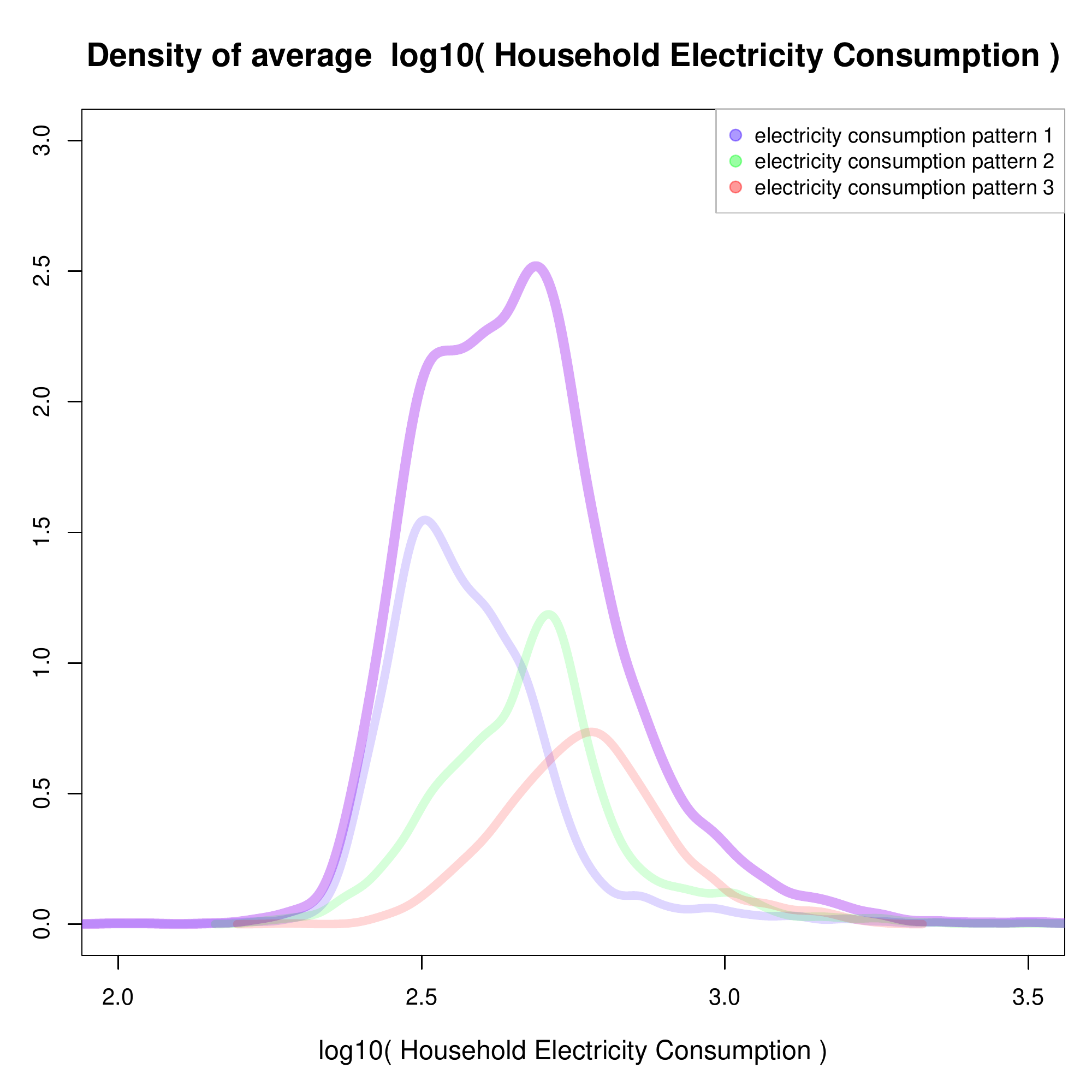}
\includegraphics[width=3.05in]{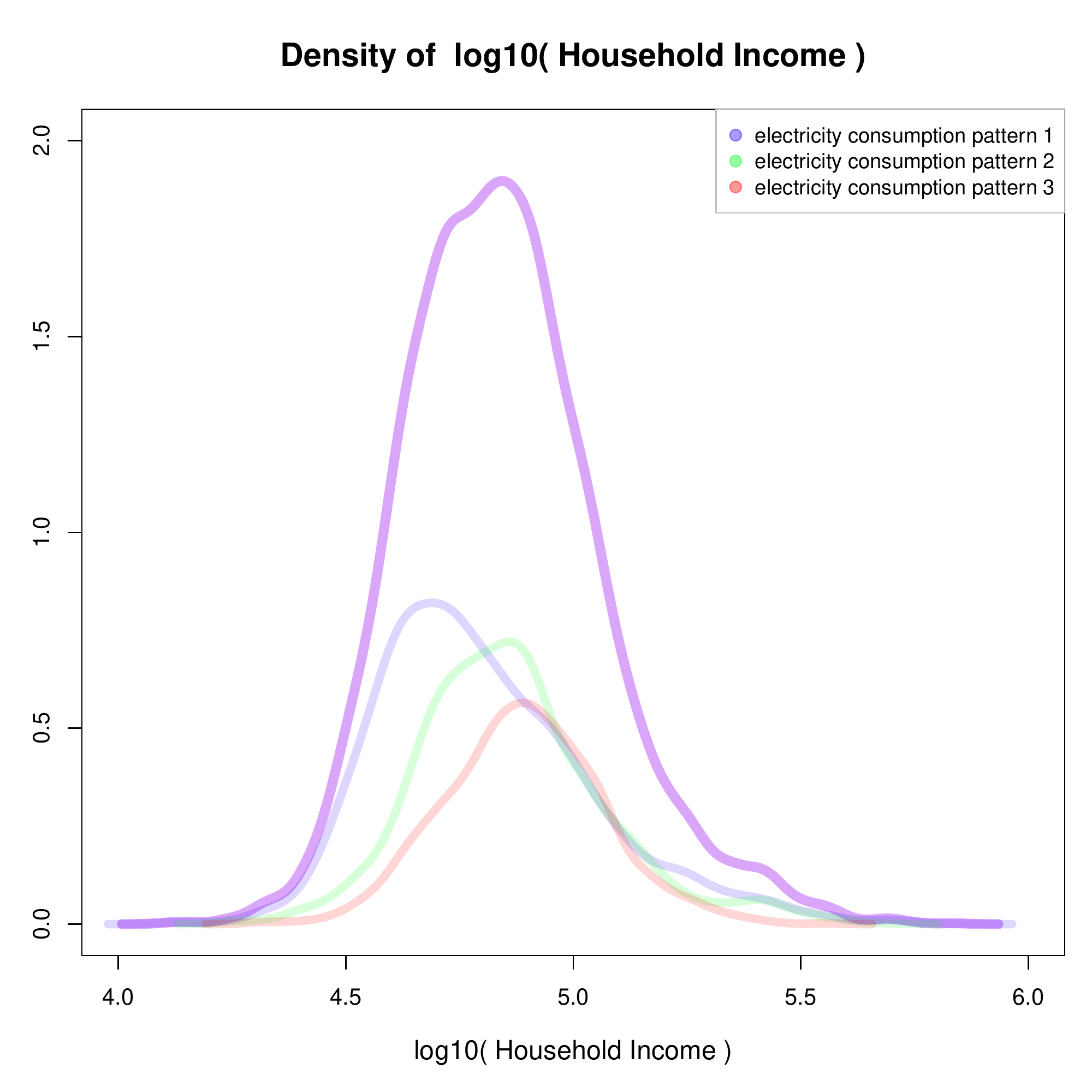}
\vspace{-0.2in}
\caption{
Density plots for log10( Household Electricity Consumption ) and log10( Household Income ).
The histograms on the left cover all values,
and the superimposed density estimates show the histograms are not quite unimodal, and not quite gaussian.
The curves on the right show the weighted density estimates obtained for each of the three clusters discussed above.
The density estimate for the overall distribution is the weighted sum of their densities.
\label{Densities}
}
\end{figure*}

A panoramic display of block group $\logHEC$ data is in Figure~\ref{logHEC_heatmap}.
Each row in this figure shows the history of one Census block group in Los Angeles County,
showing its variation in consumption over the 72 months 2006-2011.
The rows have been hierarchically clustered by similarity (of their consumption pattern);
hierarchical structure is displayed in the dendrogram on the left.
The neighborhood name for about every 50th block group is shown, suggesting that the clusters
correspond both to geographical regions and to climate in Los Angeles.

With the dataset of Figure~\ref{logHEC_heatmap} in mind, in this paper we define
an \emph{electricity consumption pattern} as a sequence of 72 monthly values (over the six years).
The patterns show clear annual cycles, with strong seasonal influences on consumption.
However, the Summer consumption pattern visible in the lower half of the heatmap
appears almost absent in the top half, where the highest consumption appears to be in Winter.
The upper half of the table shows very different consumption patterns than the lower half.
(The upper half is coastal, while the lower half is inland.)

The distribution of Household Electricity Consumption values appears approximately lognormal ---
i.e., 
the distribution of
$\logHEC$
is approximately normal.
The log-transformed distribution actually has slight positive skew,
and is sometimes modeled as a mixture of normal distributions \cite{IncomeDistributions}.

Figure~\ref{Densities} shows the distribution of $\logHEC$
over all block groups.
The histogram is not quite normal, but resembles a weighted sum (mixture) of normal distributions.

\bigskip

Household Electricity Consumption (HEC) in
these zones turns out to be strongly correlated with a number of variables,
including income, topography, and property values.
HEC patterns in each zone differ seasonally,
and are influenced by economic trends over time.
Our analysis of $\logHEC$ patterns suggests the importance of income and economic variables, as well as climate,
in the similarity and variability of annual electricity consumption patterns.

\begin{figure*}[htbp]
\centering

\includegraphics[width=6.4in]{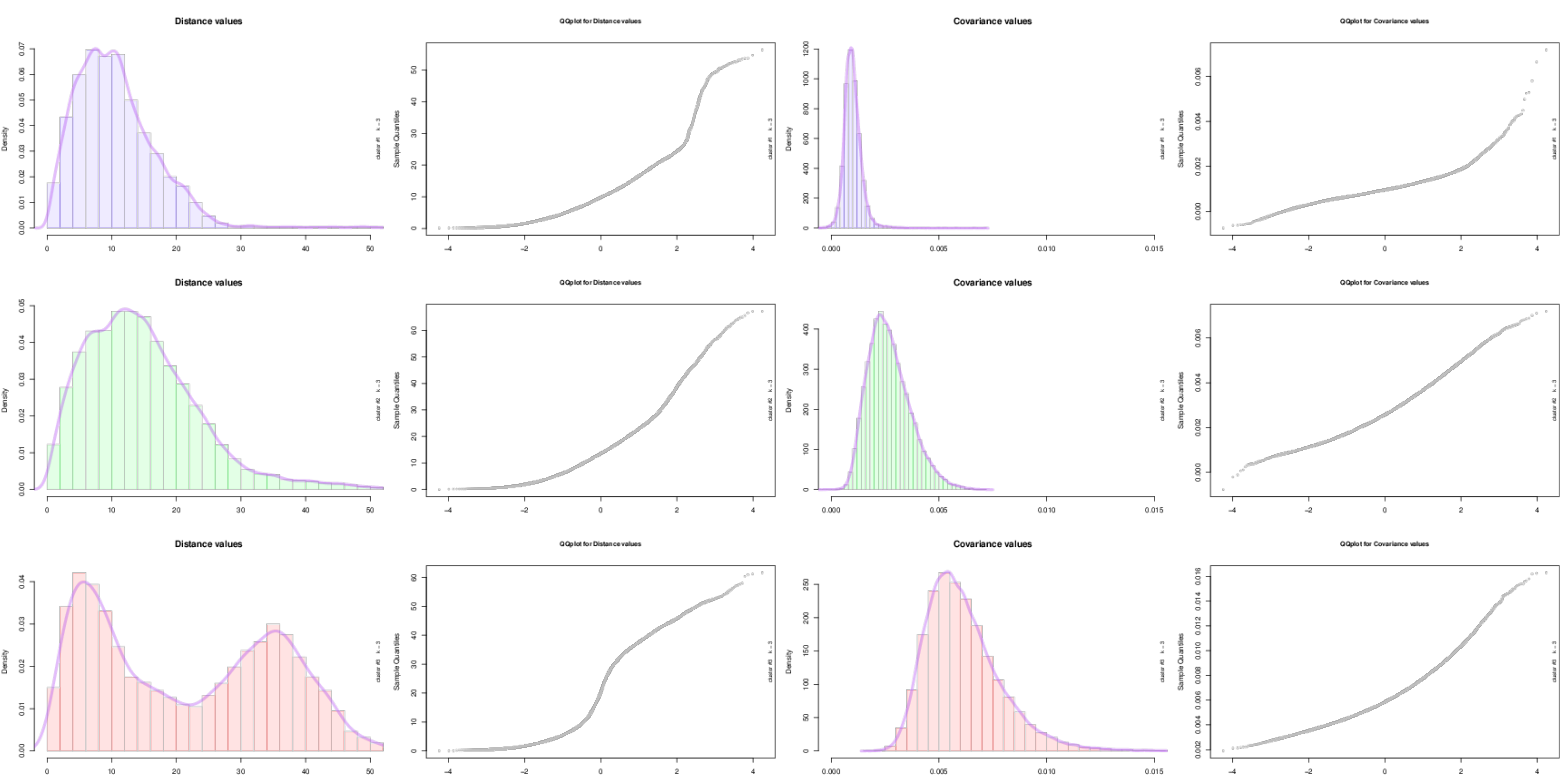}  

\caption{{
For the $\logHEC$ data, the distribution of Distance values and Covariance values appear roughly Gaussian.
However, the 12 plots show that, when the data is split into the 3 zones discussed above,
the distributions of Distance values (left) and Covariance values (right) in each zone become more nearly Gaussian,
as required by a GPRF.
In other words, $\logHEC$ more closely approximates a Gaussian Process when the data is split into
the three EC-microclimate zones.
The third zone (bottom row) shows a bi-modal distance distribution
because the region has two distinct geographic parts;
dividing the parts into two zones would eliminate the bi-modality.
Also, the distribution of (Dist,cov)-pairs resembles a 2D Gaussian, as required by a GPRF.
\label{GPValidation}
}}
\end{figure*}

\subsection{log(HEC) as a Gaussian Process Random Field}

The HEC data can be modeled as a Log-Gaussian Process over a 2D geographic space.
More precisely, when $\logHEC$ is viewed as a random variable depending on its (longitude,latitude)-coordinates,
we claim it can be modeled as a Gaussian Process Random Field.

A \emph{Gaussian Process} ${\cal GP}(m, k)$ is often defined as a collection of random variables,
any finite number of which have joint Gaussian distributions defined by a mean function $m$
and covariance function $k$ \cite{GaussianProcesses}.
This is often viewed as describing a distribution over \emph{functions},
where `a finite number of variables' corresponds to `the function values at a finite number of input points'.
Specifically,
when $\Vx$, $\Vx'$ are two points of the input space (on which $m$ and $k$ are defined as functions),
functions $f$ are said to follow this distribution
\(
f ~ \sim ~ {\cal GP}(m(\cdot), k(\cdot,\cdot))
\)
when
\begin{eqnarray*}
m(\Vx) & = & E[ ~ f(\Vx) ~ ] \\
k(\Vx,\Vx') & = & E[ ~ (f(\Vx) - m(\Vx)) ~ (f(\Vx') - m(\Vx')) ~ ] .
\end{eqnarray*}
Generally, random processes involve index values like $\Vx$ that are one-dimensional (such as `time').
When values $\Vx$ are $d$-dimensional with $d>1$, the Gaussian Process is called a
\emph{Gaussian Process Random Field}.

A common form of covariance function on pairs of random variables is the \emph{squared exponential}
\[
\mbox{cov}(f(\Vx),f(\Vx')) ~=~ k(\Vx,\Vx') ~=~ \exp(\, \mbox{\small $-\frac{1}{2}$} \, \norm{\Vx-\Vx'}^2 \,) .
\]
(In Gaussian Process regression, it is common to use instead
\[
\theta_0 ~ \exp(\, \mbox{\small $-\frac{1}{2}$}\, \theta_1 \, \norm{\Vx-\Vx'}^2 \,)  ~+~ \theta_2 ~+~ \theta_3 \, \adjoint{\Vx} \Vx'
\]
where $(\theta_0,\theta_1,\theta_2,\theta_3)$ are parameters that are fit by the regression \cite[\S6.4.2]{Bishop}.)

Another way of presenting this is to say that the $\Vx$ values are from a continuous \emph{input space},
and for each $\Vx$,
the Gaussian process $f(\Vx)$ is a normally distributed random variable.
Furthermore,
given $n$ values $(\Vx_1, \ldots, \Vx_n)$,
the $n$ random variables $(f(\Vx_1), \ldots, f(\Vx_n))$ have a $n$-dimensional normal distribution
with mean $(m(\Vx_1), \ldots, m(\Vx_n))$ and $n \times n$ covariance matrix $K$ with entries
$K_{ij} \,=\, \mbox{cov}(f(\Vx_i),f(\Vx_j)) \,=\, k(\Vx_i,\Vx_j)$.

For our HEC data, we let $\Vx$ denote a 2D (longitude, latitude)-position in degrees,
and $\Vx$ and $\Vx'$ be two such positions.
Then the \emph{geographical distance} $\distance{\Vx}{\Vx'}$
between them is their major-arc distance (when converted to radians), i.e., 
spatial distance, in miles.
Thus $\distance{\Vx}{\Vx'}$ is approximately proportional to $\norm{\Vx-\Vx'}^2$.

Also, at each such 2D position $\Vx$, we have a random variable $f(\Vx) = \log(\mbox{HEC}(\Vx))$.
For the block group located at $\Vx$,
our dataset contains 72 samples of this random variable, a time series of monthly values over 2006-2011.
Treating the samples as vectors, $\mbox{cov}(f(\Vx),f(\Vx'))$ can be estimated as the vectors' covariance.

\medskip

We validate these properties below after defining EC-microclimate zones more formally.

\section{Seasonal Electricity Consumption Patterns and Microclimate Zones}

\subsection{Clustering of EC Patterns}

Figure~\ref{basic_electricity_consumption_map} shows three average electricity consumption patterns
obtained by clustering, motivated by Figure~\ref{logHEC_heatmap}.
The EC patterns of each block group
(a sequence of length 72 $\logHEC$ values, for each month over the six years 2006-2011,
normalized to total to 1)
were clustered using the $k$-means algorithm into $k$ = 3 clusters.
To visualize the result,
a single point was then plotted at the central (latitude,longitude) location of the block group,
and colored according to its cluster.
The map shows a connection between electricity consumption and geography,
suggesting association between EC and climate.

The electricity consumption patterns of each block group were first computed
as a sequence of 72 values of $\logHEC = \log_{10}(\mbox{EC per household})$,
for each month over the six years 2006-2011, that were also normalized to total to 1.
These sequences were clustered using the $k$-means algorithm
(in R, with standard settings and 25 random restarts)
into $k = 3$ clusters.

The log-transformation is natural (and matters) for HEC patterns.
As mentioned earlier,
the distribution of $\logHEC$ is approximately normal,
and the distribution of HEC is highly skewed.
Second, numeric measures like HEC are often multiplicative in nature, the product of many independent factors,
and so their log is a sum of independent factors (and ultimately: normal).
Economic variables that range over multiple orders of magnitude also have this property
\cite{Energy_Forecast_with_lists_of_Variables}.

The normalization of HEC patterns to total to 1 also matters,
as it makes their shape significant, not their scale.
If HEC patterns reflect climate variation,
and we cluster them by Euclidean distance (as in $k$-means clustering),
then the absolute scale of the patterns (scale of consumption) can be factored out; normalization removes this scale.
Later on Principal Component Analysis also reveals that patterns in a climate zone differ primarily by scale multipliers.
So both log-transformation and scale normalization improve results of clustering.

Other distance measures could have been used in clustering.
Correlation of HEC patterns is a natural alternative, for example, and time-series measures of similarity also.
Euclidean distance may not be the best measure, but we show later with PCA that annual energy use patterns in each cluster
actually have limited variation.
Normalizing the total to 1 factors out the scale of the patterns, which turns out to be their primary feature.
Normalized patterns can then be compared with Euclidean distance.

\subsection{EC-Microclimates}

A \emph{Microclimate} is often defined by a restricted range of environmental variables,
including temperature, wind, relative humidity, and vegetation \cite{Microclimates}.
Cities (such as Los Angeles) with diverse patterns of variables
support diverse microclimates.

\medskip

The regions defined by clusters visible in Figure~\ref{basic_electricity_consumption_map}
are contiguous geographic regions,
although they are defined only by EC patterns.
Furthermore, they behave like climate zones, reflecting environmental characteristics
that impact electricity consumption over the year.
The three regions in the figure --- western Los Angeles, the San Gabriel valley, and the San Fernando valley ---
all have distinct microclimates.

\medskip

For this reason we refer to them as \emph{EC-Microclimate Zones}.

\medskip

It is well-known that microclimates affect energy use \cite{UrbanPhysics},
but we have not found any discussion of the reverse,
i.e., that EC is sufficient to identify microclimate.

\medskip

This is the central finding of our work:
\emph{EC-microclimate zones}
--- regions in which block groups have similar Electricity Consumption patterns over time ---
resemble environmental microclimate zones. 
HEC data includes a history of climate.

\section{Validity of EC-Microclimate Zones as Microclimate Zones}

Environmental microclimates appear to cause predictable seasonal EC patterns,
and different microclimates have patterns that differ in characteristic ways.
Analogously, we can show that EC-microclimates resemble environmental microclimates in several ways:
(1) both environmental- and EC-microclimates exhibit random field structure;
(2) both conform to topography;
(3) the clusters of EC patterns in EC-microclimates exhibit limited variance.
In these perspectives, EC-microclimates behave like actual microclimates.

\subsection{Microclimates and EC-Microclimates as Random Fields}

It has been observed before that microclimate variables
(wind speed, local temperature, wind pressure, and total solar irradiation)
can be modeled as a Gaussian Process Random Field,
and this has been exploited in modeling electricity consumption
\cite{UncertaintyOfMicroclimateVariables}.

In this paper we do the reverse:
show that electricity consumption in each zone apparently can be modeled as a random field
($f = \logHEC$ approximates a Gaussian Process).
The result is novel, yet consistent with \cite{UncertaintyOfMicroclimateVariables}.
Both microclimates and EC-microclimates can be modeled by similar random fields.
In each zone, $\logHEC$ is a summary of microclimate variables.
So given two distinct microclimates,
we expect the $\logHEC$ patterns of corresponding EC-microclimate zones to be distinct.
Figure~\ref{basic_electricity_consumption_plots} shows that this does hold for our data.

\subsection{Topography, Basins, Valleys, and Microclimates}

The EC-microclimate zones in Figure~\ref{basic_electricity_consumption_map}
that are bounded by hillsides are essentially basins, each with a distinct climate.
Figure~\ref{HouseholdIncomeMap} shows corresponding Los Angeles topography.

The red region to the northwest is an inland area (the San Fernando Valley),
while the blue regions to the southwest are coastal areas (western Los Angeles).
These two regions have different climates.
The adjacent green region (eastern Los Angeles and the San Gabriel Valley)
is surrounded by hills, with a warm climate.
In this perspective, clustering the EC patterns has managed to identify regions separated by topography,
because their characteristic patterns are distinct.

Generally, EC-microclimate zones we have produced conform to basin-like regions in Los Angeles.
In \cite{Micromodels} and \cite{Micropolicy}, we show that increasing the value of $k$
identifies a more refined set of regions, again with topographical features separating them.
Below, we show that limited variance of EC patterns in each zone offers an explanation for this structure.

\subsection{Properties of k-Means Solutions}

We have used $k$-means clustering to obtain microclimate zones for small values of $k$ (up to 10).
The $k$-means algorithm performs greedy minimization of the total within-cluster sum of squares
\(
\mbox{WSS}(k) \,=\, \sum_{i=1}^k \, \sum_{j \in C_i} \, \mid\mid\Vv_j - \vect{\mu}_i \mid\mid^2
\)
\cite{Murphy} for $k$ clusters,
where the $C_i$ ($1 \leq i \leq k$) are clusters with mean $\vect{\mu}_i$, containing points $\Vv_j$
(vectors of 72 $\logHEC$ values, normalized so that they sum to 1).
A natural question is whether this clustering algorithm has some justification.

First, the stability of the clusters obtained is a central question.
The number of random restarts used with $k$-means is important here: 
if the clustering ultimately returned is the best clustering obtained from these restarts,
then higher numbers of restarts give more stable results.
In other words, the clusters returned by $k$-means are identical more often with more restarts.

Figure~\ref{StabilityAnalysis} shows that, for our data with 100 random restarts, clusters for $k \leq 5$ were stable.
More specifically, as nstarts (the number of restarts) increases, the number of points (block groups) with more than one

Beyond this, some data points can lie near the boundary of two clusters, and the randomness in $k$-means can assign them to one or the other.
With 100 restarts, there are few such points with $k \leq 8$ clusters.
Even higher numbers of restarts can reduce the number, but because $k$-means is greedy, in different runs might yield clusters that differ on a few points.
It is important to realize that the randomness in $k$-means could yield clusters that vary in this way.

\begin{figure*}
\scriptsize
\centering
\begin{tabular}{|r|rcr|rcr|rcr|rcr|rcr|}
\hline
$k$ & nstart & = & 10 & nstart & = & 25 & nstart & = & 50 & nstart & = & 100 & nstart & = & 250 \\
\hline
 & & & & & & & & & & & & & & &   \\[-4pt]
2 & 0 / 5474  & = &   0.0 \% & 0 / 5474  & = &   0.0 \% & 0 / 5474  & = &   0.0 \% & 0 / 5474  & = &   0.0 \% & 0 / 5474  & = &   0.0 \% \\
3 & 0 / 5474  & = &   0.0 \% & 0 / 5474  & = &   0.0 \% & 0 / 5474  & = &   0.0 \% & 0 / 5474  & = &   0.0 \% & 0 / 5474  & = &   0.0 \% \\
4 & 0 / 5474  & = &   0.0 \% & 0 / 5474  & = &   0.0 \% & 0 / 5474  & = &   0.0 \% & 0 / 5474  & = &   0.0 \% & 0 / 5474  & = &   0.0 \% \\
5 & 5 / 5474  & = &   0.1 \% & 5 / 5474  & = &   0.1 \% & 0 / 5474  & = &   0.0 \% & 0 / 5474  & = &   0.0 \% & 0 / 5474  & = &   0.0 \% \\
6 & 2970 / 5474  & = &  54.3 \% & 29 / 5474  & = &   0.5 \% & 35 / 5474  & = &   0.6 \% & 35 / 5474  & = &   0.6 \% & 13 / 5474  & = &   0.2 \% \\
7 & 238 / 5474  & = &   4.3 \% & 258 / 5474  & = &   4.7 \% & 126 / 5474  & = &   2.3 \% & 2 / 5474  & = &   0.0 \% & 2 / 5474  & = &   0.0 \% \\
8 & 4242 / 5474  & = &  77.5 \% & 3 / 5474  & = &   0.1 \% & 3 / 5474  & = &   0.1 \% & 0 / 5474  & = &   0.0 \% & 0 / 5474  & = &   0.0 \% \\
9 & 1826 / 5474  & = &  33.4 \% & 288 / 5474  & = &   5.3 \% & 282 / 5474  & = &   5.2 \% & 35 / 5474  & = &   0.6 \% & 52 / 5474  & = &   0.9 \% \\
10 & 3509 / 5474  & = &  64.1 \% & 1632 / 5474  & = &  29.8 \% & 42 / 5474  & = &   0.8 \% & 287 / 5474  & = &   5.2 \% & 11 / 5474  & = &   0.2 \% \\
11 & 3644 / 5474  & = &  66.6 \% & 15 / 5474  & = &   0.3 \% & 9 / 5474  & = &   0.2 \% & 5 / 5474  & = &   0.1 \% & 0 / 5474  & = &   0.0 \% \\
12 & 5409 / 5474  & = &  98.8 \% & 1835 / 5474  & = &  33.5 \% & 1844 / 5474  & = &  33.7 \% & 1261 / 5474  & = &  23.0 \% & 37 / 5474  & = &   0.7 \% \\[2pt]
\hline
\end{tabular}
\caption{Table showing the number of block groups (the 5000 points in the plots) that k-means assigns to more than 1 cluster (and corresponding percentage).
Points whose clusters change are usually located around the boundary between two clusters, and therefore k-means can assign it to either one.
(Each entry in the table reflects the percentage of points that changed over 10 runs of k-means, with the indicated values of $k$ and nstart.)
In our implementation of k-means, the value of nstarts (= number of random restarts) influences the optimality of the clustering:
Random restarts are used to reduce the chance of its k-means' greedy search becoming trapped in a local minimum.
As $k$ gets larger, more random restarts are needed to obtain clusters that are consistent.
The table shows that large values of $k$ (such as 11 or 12) can yield stable clusters if a sufficiently large value of nstart (such as 250) is used.
\label{StabilityAnalysis}
}
\end{figure*}

Optimizing the $\mbox{WSS}(k)$ objective function obtains spherical clusters around the $k$ centroids
$\vect{\mu}_i$ (the $k$ mean $\logHEC$ patterns).
Basins and valleys do often have rough spherical shape, and 
$k$-means appears to do well when zones correspond to basins ---
a situation in which they are justifiable as defining microclimates.
In \cite{Micromodels} we explore optimization of $k$; in this paper we only study results for $k=3$.

Instead of using a clustering algorithm, we could use a method for extracting \emph{mixture models} \cite{Murphy},
and obtain probability distributions over models rather than separate clusters.
We have elected to use clustering because boundaries (and policy) of municipalities are \emph{hard}, preventing mixtures.
All zones in this paper are regions having a single model, not a mixture.

\medskip

In each of these three perspectives ---
random fields, topography, and results of clustering ---
EC-Microclimates behave like microclimates.

\begin{figure*}[htbp]
\centering
\vspace{-0.25in}
\includegraphics[width=7.25in]{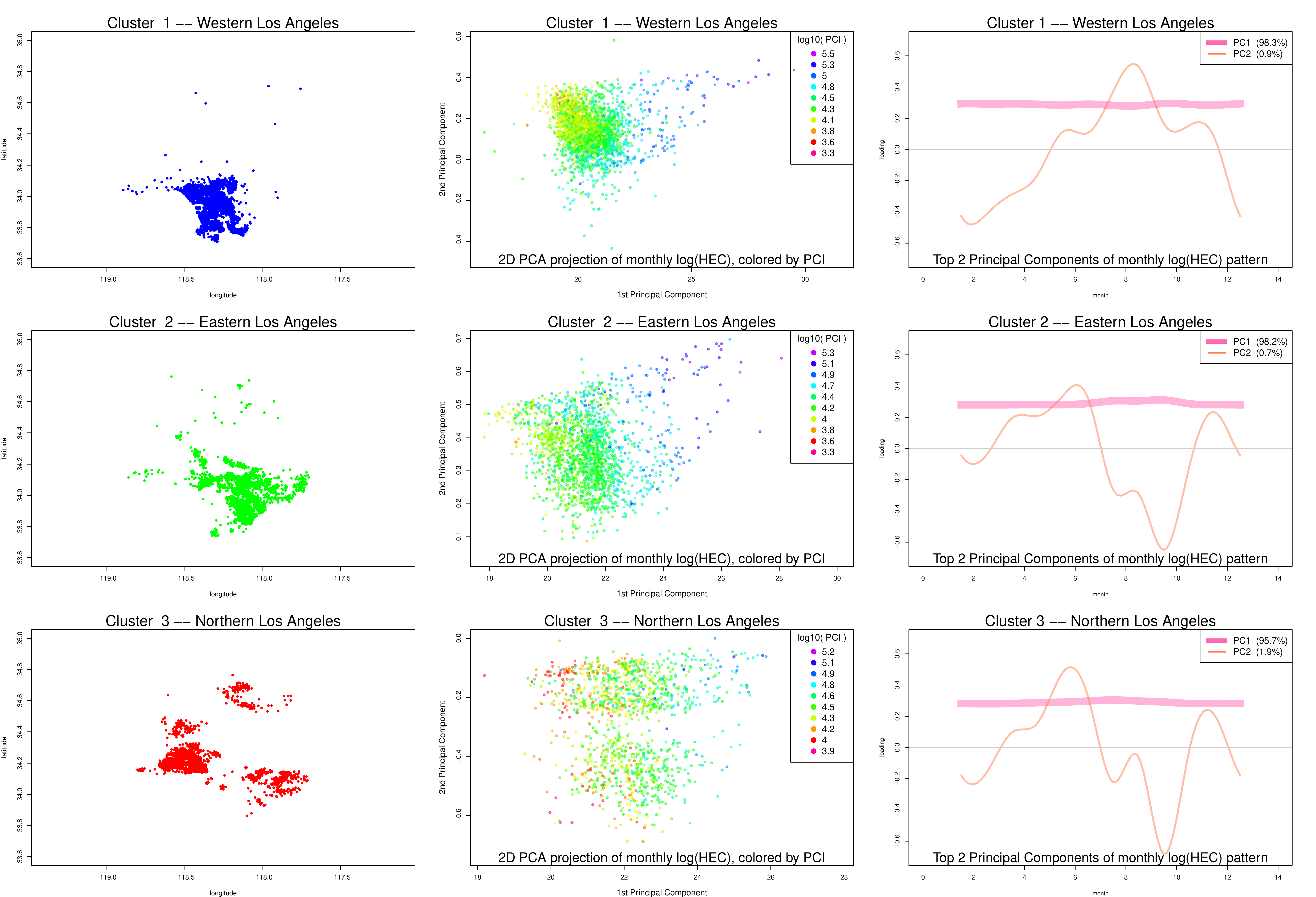}
\vspace{-0.1in}
\caption{
The plots show different aspects of Principal Components Analysis \cite{Bishop} on 12 \emph{monthly} averages of $\logHEC$
for about 5000 block groups in Los Angeles County.
The data has been divided into the three zones derived earlier,
and each row shows results for average annual $\logHEC$ patterns in one zone.
(The sequences of 12 values are not normalized to sum to 1; they reflect magnitude of consumption.)
The second column shows each zone's principal component structure;
the coloring of points indicates the log of Per Capita Income (PCI); points with higher PCI appear more blue.
This coloring shows that the first two principal components directly reflect PCI, which is a correlate of Property Values.
The third column shows the first 2 principal components for each zone.
The first component, which explains more than 95\% of the variance, are the wide nearly-horizontal patterns.
The second component is an annual curve summarizing most of the remaining variance.
\label{MEC_by_Cluster}
}
\end{figure*}

\section{Electricity Consumption Patterns \\ and Per Capita Income}

Electricity consumption is often linked to economics.
For example, in \cite[p.65]{ElectricPowerDistribution}
electricity consumption in the United States is linked to GNP.
Furthermore it is known that individuals with the greatest income
have the greatest consumption \cite{HouseholdConsumption},
while lifeline consumers have much lower use \cite{JackiMurdock}.

In this section we develop simple econometric models relating HEC to income --- like crude HEC forecasting models.
Our results above show seasonal consumption is linked to microclimate.
Income distributions often resemble
a mixture of lognormal distributions \cite{IncomeDistributions}.
Figure~\ref{Densities} shows that this holds for the distribution of $\logPHI$
as well as the distribution of $\logHEC$.

\subsection{Linking Household Income and Electricity Consumption}

The map of Household Income in Figure~\ref{HouseholdIncomeMap} shows that
higher-income block groups are located in the western Los Angeles basin,
particularly in hillside and coastal areas,
while lower-income block groups are mainly in flat inland areas.
The microclimates of these locations differ.

The association between microclimate and income helps explain why higher-income consumers
have different electricity consumption profiles.
Figure~\ref{basic_electricity_consumption_plots}
suggests that higher-income block groups have higher consumption in Winter,
while lower-income block groups have higher consumption in Summer.
The 3 microclimates have different seasonal demand.

\begin{figure}[htbp]
\centering
\includegraphics[width=3.2in]{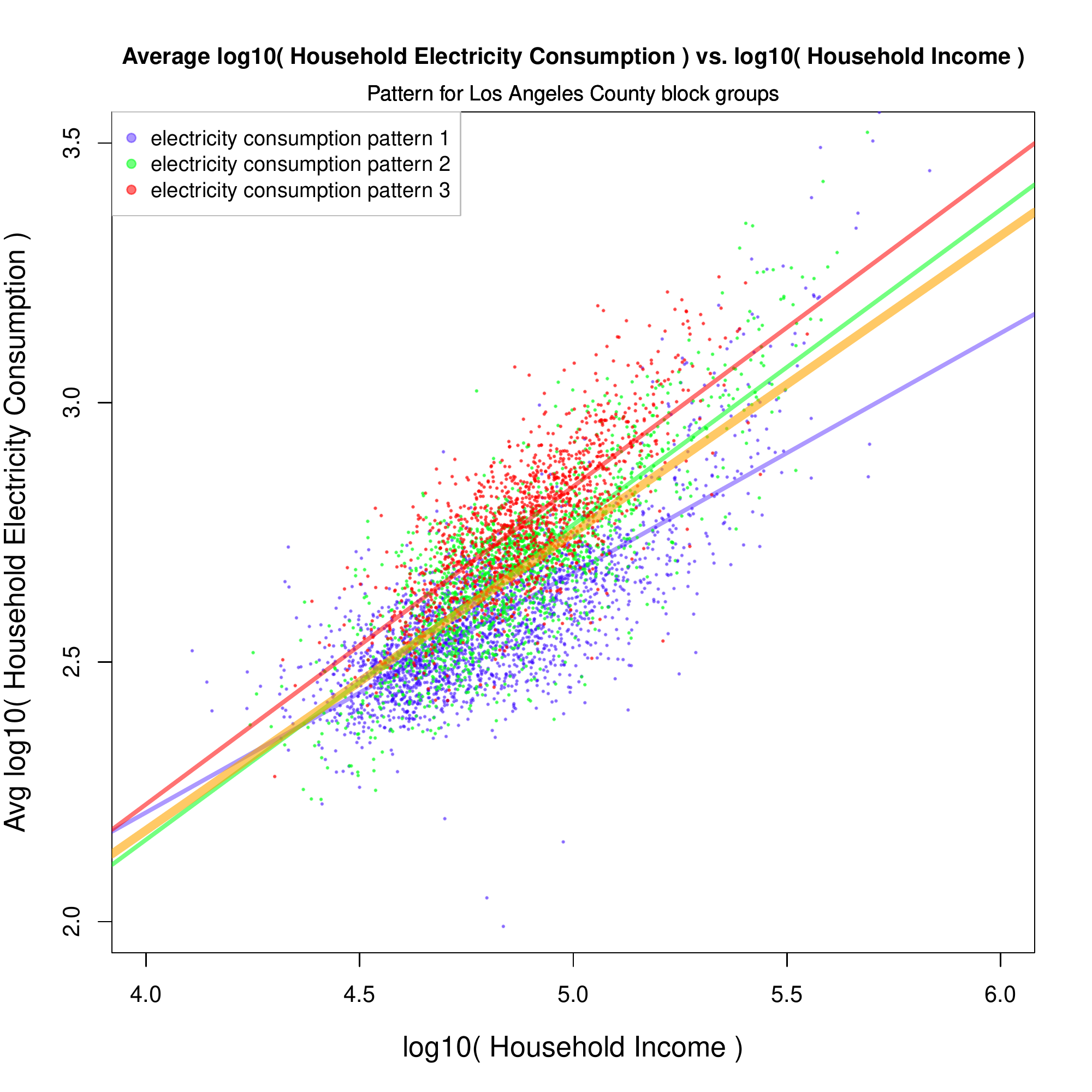} \\[-10pt]

{\footnotesize
\begin{eqnarray*}
\logHEC  \sim ~ 0.57 \logPHI - 0.11 & R^2 = 0.56 & \mbox{(overall)} \\
\hline
& & \\[-3pt]
\logHEC  \sim ~ 0.46 \, \logPHI + 0.36 & R^2 = 0.57 & \mbox{(cluster 1)} \\
\logHEC  \sim ~ 0.60 \, \logPHI - 0.27 & R^2 = 0.64 & \mbox{(cluster 2)} \\
\logHEC  \sim ~ 0.61 \, \logPHI - 0.22 & R^2 = 0.57 & \mbox{(cluster 3)}
\end{eqnarray*}
}
\vspace{-0.25in}
\caption{
Average Household Electricity Consumption for each of about 5000 block groups in Los Angeles County,
versus Per Capita Income.
This plot also shows the result of \emph{clustered regression} -- i.e., shows a regression model for each cluster;
all \emph{log}s are base 10.
Higher-income consumers appear more to the right,
and higher-consumption consumers appear higher in the display.
The western coastal zone appears below the regression line (higher income, lower consumption)
and the land-locked northern zone appears above it (lower income, higher consumption).
Three lines show linear regression models for each of the 3 zones;
the wider orange line is a model for all of the data.
\label{ClusterRegression}
}
\end{figure}

\subsection{Variation of Electricity Consumption and Income}

To understand the magnitude of seasonal variations of HEC for the block groups,
we can use Principal Components Analysis (PCA) \cite{Bishop}.
Figure~\ref{MEC_by_Cluster} shows the result of PCA
on sequences of 12 monthly averages of $\logHEC$ over the six years.
Points are colored by Per Capita Income (PCI) of the regions.

One interesting aspect of the resulting components is that they give a straightforward model of
annual electricity consumption that explains 99+\% of the variance in annual electricity consumption with three components,
and the first alone explains 95.7\%.

\medskip

The plots at right in Figure~\ref{MEC_by_Cluster}
show the first 2 principal components presented as consumption patterns over the twelve months.
The first component is an `augmentation' above the baseline that is almost evenly-spread across all 12 months;
the other components depend on the zone.
The first component explains almost all of the variance.
Essentially, all patterns in a microclimate differ almost entirely by a small additive amount
--- given by the coefficient of the first principal component.
Since the first component is nearly constant, the difference is nearly an additive constant.

\medskip

In Figure~\ref{MEC_by_Cluster},
the plots in the middle show the block groups projected on the first 2 principal components.
There is a clear color gradient from blue to red in both dimensions;
block groups with higher PCI generally have higher coefficients in all components.
Thus: regional variation in electricity consumption is \emph{strongly} linked to income.
Because PCI is a correlate (proxy) of property values,
this also shows that higher property values are associated with greater variation in HEC.

\begin{figure}[htbp]
\centering
\includegraphics[width=3.2in]{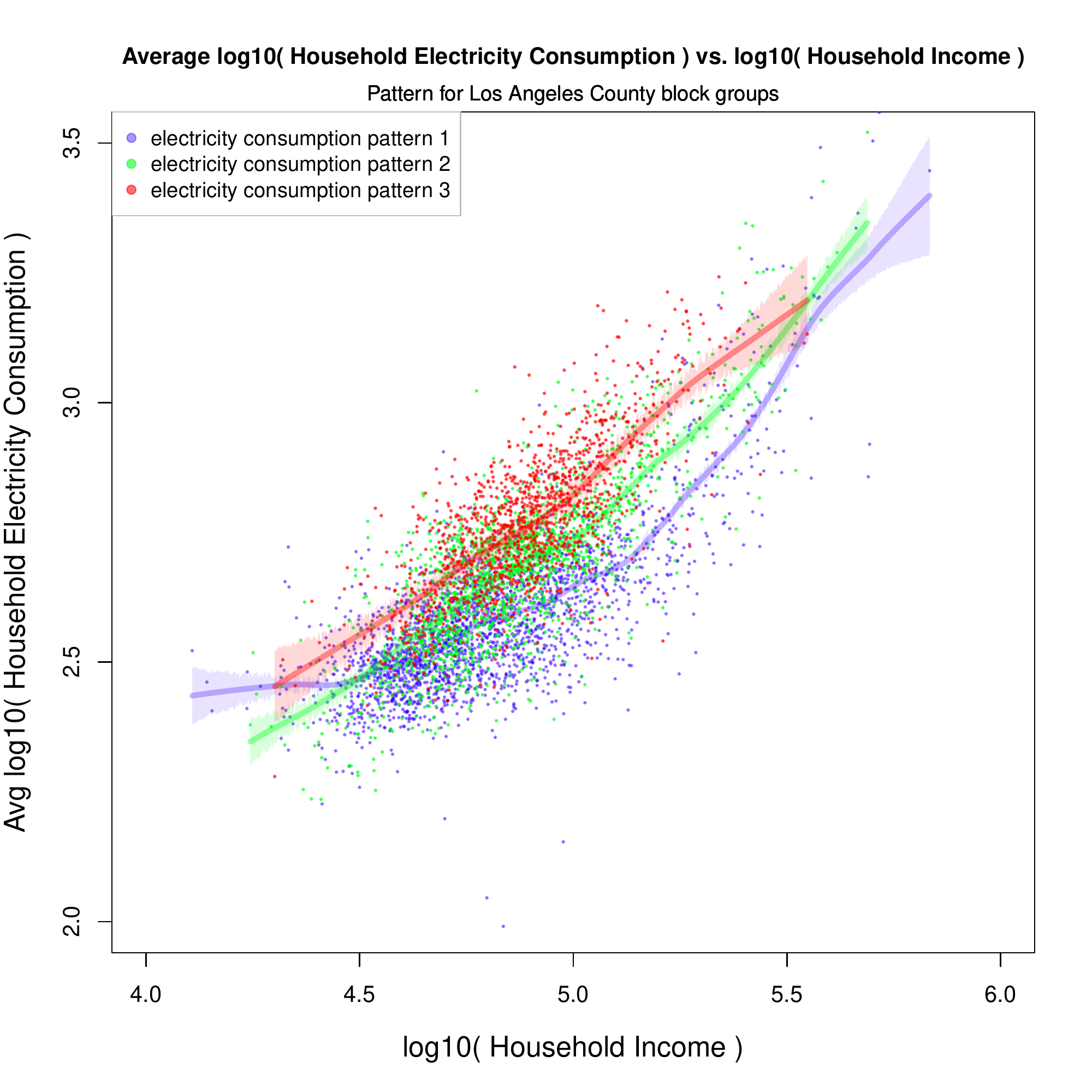}
\vspace{-0.1in}

\caption{{
This plot shows results of \emph{clustered Gaussian Process Regression}
for Average Household Electricity Consumption ($\logHEC$)
versus Household Income ($\logPHI$)
--- with one model for each cluster.
The three clusters are the same as in Figure \ref{ClusterRegression}.
The shaded $\pm 2$-standard-error bands are much wider at high or low values of PHI.
Gaussian Process models fit the data more closely than linear models,
and can provide more accurate prediction \cite{GPprediction}.
\label{GPRegression}
\label{ClusterGPRegression}
}}
\end{figure}

\medskip

How can one principal component explain so much of the variance?
Notice that the principal component information tells us a great deal about each zone.
Any given 12-month $\logHEC$ pattern $\vect{v}$ in Zone $i$ satisfies
$\vect{v} \;\approx\; \vect{\mu}_i \,+\, c\, \vect{e}_1$,
where $\vect{e}_1$ is the first principal component,
and $c = ((\vect{v}-\vect{mu}_i)' \, \vect{e}_1)$.
But since the first principal component is nearly constant,
$\vect{e}_1 \;\approx\; \vect{1}/\sqrt{m}$, and
$\vect{v} \;\approx\; \vect{\mu}_i \,+\, c \, \vect{1}/m$.
In other words, every vector in zone $i$ is essentially $\vect{\mu}_i$ plus an additive constant `shift' determined by $c$.

Because adding a constant determined by $c$ to $\logHEC$ is equivalent to multiplying HEC by a scaling factor, 
higher-income block groups differ from others in the zone only in having a higher scale.

\subsection{Electricity Consumption in each Microclimate Zone}

In our data, division into microclimate zones (i.e., conditioning on EC-microclimate)
yields more accurate models of electricity consumption.
Figure~\ref{MEC_by_Cluster} shows the same result when the data is subdivided into the three clusters identified earlier
(western, eastern, and northern Los Angeles).
As one would expect, this division permits even more accurate results from PCA.
Again, however, a single principal component explains almost all variance in each cluster/region.

Figure~\ref{ClusterRegression} shows the joint distribution between
$\log(\,\mbox{PHI}\,)$ and $\logHEC$.
The distribution appears almost gaussian, with strong correlation.
Subdividing the block groups over the three clusters
gets three smaller gaussian-like distributions, with stronger correlation.
Electricity consumption patterns differ visibly in each EC-microclimate zone.

\subsection{Predicting Electricity Consumption}

Distributions of $\logHEC$ values often appear gaussian,
particularly within EC-microclimate zones.
Earlier we discussed the Gaussian Process structure of the $\logHEC$ data;
in fact the structure depends on the zone.

Figure~\ref{logHEC_heatmap} shows how the values of $\logHEC$ differ by zone;
each row is a time series of 72 samples for a given block group.
At least 3 different kinds of patterns are visible,
and were detected as EC-microclimate zones by clustering.
Thus the $\logHEC$ random field discussed earlier
is better represented by 3 different GPRFs, one for each zone.

\bigskip

Gaussian Process Regression \cite{GaussianProcesses} is a method for learning a Gaussian Process
from a given set of data points.
For a training set $(X,\Vy) = \{(\Vx_i,y_i) \mid 1\leq i \leq n\}$
defining a function $f$,
it derives a covariance matrix $K = K(X,X)$ that can be used to make predictions for the function value
at any new input point $\Vx_*$:
\[
\overline{f}(\Vx)
~=~ \adjoint{\Vk(\Vx_*)} (K + \sigma_n^2)^{-1} \Vy ,
\]
where $\Vk(\Vx_*) = K(X,\Vx_*)$,
and $\mbox{cov}(\Vy) = K + \sigma_n^2 I$ 
(i.e., $\sigma_n^2$ is the level of noise in the $\Vy$ values \cite{GaussianProcesses,Bishop}).

Figure~\ref{ClusterGPRegression} shows the result of \emph{Clustered Gaussian Process Regression},
performing Gaussian Process regression for $(\logPHI,\logHEC)$ points in each cluster (EC-microclimate zone).
Again, clustering the data improves accuracy significantly.
These results give a better fit to the data than the linear regression models in Figure~\ref{ClusterRegression};
they also show that household electricity consumption increases nonlinearly
for households of extremely high income.
The highest-income block groups appear price-insensitive.

\medskip

\emph{Gaussian Process Prediction}, or \emph{Kriging}, is the use of
Gaussian Process Regression to obtain predicted values \cite{GPprediction}.
The regression can yield predictions for any input $\Vx$.
Their accuracy depends on noise in the data, but Gaussian Process regression also produces standard error bounds,
providing a quantitative estimate of accuracy.
In Figure~\ref{ClusterGPRegression}, each regression line includes $\pm 2$-standard-error `band';
these bands are tallest for extreme values of Household Income (PHI), high or low.
The plot shows the variance is heteroskedastic, and is greatest at high income:
EC of the wealthiest neighborhoods is most difficult to predict accurately.

\medskip

Kriging is an important method of geostatistics,
used often to make predictions based on spatio-temporal data \cite{GaussianProcesses}.
It might have potential in prediction of energy demand.

\newcommand{\logPPH}{{\log(\,\mbox{PPH}\,)}}
\newcommand{\logeHEC}{{\log_{e}(\,\mbox{HEC}\,)}}
\newcommand{\logePHI}{{\log_{e}(\,\mbox{PHI}\,)}}
\newcommand{\logePCI}{{\log_{e}(\,\mbox{PCI}\,)}}
\newcommand{\logePPH}{{\log_{e}(\,\mbox{PPH}\,)}}
\newcommand{\logtHEC}{{\log_{10}(\,\mbox{HEC}\,)}}
\newcommand{\logtPHI}{{\log_{10}(\,\mbox{PHI}\,)}}
\newcommand{\logtPCI}{{\log_{10}(\,\mbox{PCI}\,)}}

\def\cents{\makebox[0cm][l]{\rule{.47ex}{0mm}\rule[-.3ex]{.15mm}{1.6ex}}c}

\section{Modeling Electricity Consumption with Clustered Regression and Mixture Regression}
\label{ClusteredModels}

A 2014 prediction model for household EC developed by the state of California
is shown in Figure~\ref{CEC_model} ---
the first econometric model in \cite{Energy_Forecast_with_lists_of_Variables}.
This statewide model can be compared with the models we consider below.
Notice that this model is an aggregate model over 1980-2013,
where the models below are over the 6-year period 2006-2011.
Notice also that the model uses natural log transformations on income variables, where we have used the base-10 log.
Assuming its variable \emph{PPH = Persons Per Household} is the ratio of income Per Household to income Per Person,
\(
\logPPH \, = \, \logPHI \,-\, \logPCI .
\)
Thus this model is of the same kind as the models presented in this paper,
emphasizing per household modeling because the EC data provides that.
We could get more accurate results with separate models for each microclimate zone.

The clustering phases in microclimate zone derivation and in clustered regression can be the same.
Micro-models are models defined on a specific zone.
In this way we can integrate micro-models with microclimates.

\begin{figure}[htbp]
\begin{center}
\footnotesize
\begin{tabular}{|rcll|}
\hline
 & & & \\[-4pt]
$\logHEC$  $\sim$\hspace{-0.6in} & & &  {natural log of EC per household} \\[-1pt]
                  & & &  {by planning area, 1980-2013} \\
 + & 7.1881 & & overall constant \\
 + & 0.3935 & $\logPPH$ & Persons per Household \\
 + & 0.1419 & $\logPCI$ & Per capita income (2013\$) \\
 $-$ & 0.0042 & UnEmpRate &  Unemployment Rate \\
 $-$ & 0.0870 & ResElecRate &  Residential Electricity Rate (2013 $\cents$/kWh)\hspace{-0.075in} \\
 + & 0.0323 & log(CoolDays) &  Number of Cooling Degree Days (70$^o$) \\
 + & 0.0181 & log(HeatDays) &  Number of Heating Degree Days (60$^o$) \\
 $-$ & 0.5784 & LADWP &  Los Angeles Dept of Water \& Power \\[-5pt]
 \vdots & & &  \\[3pt]
\hline
\end{tabular}
\end{center}

{\small
\begin{quote}
The \emph{Residential Sector Electricity Econometric Model} (for \emph{California Energy Demand Updated Forecast}, 2014)
is the first model in \cite[p.A-1]{Energy_Forecast_with_lists_of_Variables}:
All coefficients are statistically significant (their estimated error is outside 3 standard deviations).
Some variables have been omitted.
The model is described as:  ``All variables in logged form except time and unemployment rate''.  In other words,
the model uses natural log transformations on variables.

This is a statewide model for California.
The `LADWP' variable is one of five utility `planning area' dummy variables,
indicating differences for the Los Angeles region covered by LADWP.
It is like the zone conditionals $Z$ in our models, except that it only adds a constant offset.
\end{quote}
}
\caption{
A \emph{Residential Sector Electricity Econometric Model} for $\logHEC$
developed by the CEC Demand Analysis Office in 2014 \cite{Energy_Forecast_with_lists_of_Variables}.
\label{CEC_model}
}
\end{figure}

\subsection{Clustered Regression models of Electricity Consumption}

Figure~\ref{MEC_by_Cluster} suggests that
division by cluster permits $\logHEC$ 
to be well-described at the block group level
with a linear statistical model involving Per Capita Income.
In the format of a general linear model \cite{ChambersHastie} conditioned on EC-microclimate zones $Z$: 
\begin{eqnarray*}
\lefteqn{\log(\,\mbox{Household Electricity Consumption}\,)} ~~ \\
& ~~\sim~~ & \log(\,\mbox{Household Income}\,) ~\mid~ Z.
\end{eqnarray*}
In other words, we obtain a model for each microclimate zone $Z$;
all logs are to base 10.
This is precisely what is shown in
Figure~\ref{ClusterRegression}.
The regression model
\(
\logHEC ~ \sim ~ 0.57 ~ \logPHI - 0.11 
\)
for the entire dataset (all three zones) with $R^2 = 0.56$,
can be split into the three models shown in Figure~\ref{ClusterRegression}
with higher $R^2$ values.
In \cite{Micromodels} we develop methods to split optimally.

\medskip

Modeling in this way has been referred to as \emph{Clustered Regression} ---
defining a set of independent regression models on subsets of the input that are defined by clusters \cite{Torgo,ClusteredMultipleRegression}.
Strategies for computing these models range from simple $k$-means clustering followed by construction of
regression models, to `$k$-regression' models that extend $k$-means clustering to an iterative regression algorithm.

\medskip

It is pointed out in \cite{Torgo} that clustered regression models are inequivalent to standard regression
trees or ensemble methods, and can outperform them significantly.
The 3 seasonal patterns (defined earlier in Figure~\ref{basic_electricity_consumption_map} and
Figure~\ref{MEC_by_Cluster}) show how the cyclic seasonal pattern is modeled more accurately
with a different regression model in each of the 3 clusters.
The $R^2$ values suggest this also, but visual confirmation is in 
Figure~\ref{ClusterRegression}:
the heavier line is a model for the whole dataset,
and the three other lines are models for the clusters.
The overall joint distribution of $\logHEC$ and $\logPHI$ appears gaussian,
but divides clearly into distinct distributions by cluster.

\subsection{Clustered vs. Mixture Models}

Figure~\ref{Densities} showed earlier that the distributions both of
$\logHEC$
and of
$\logPHI$
are approximately gaussian,
with approximately gaussian contributions from each EC-microclimate zone.
As a result they are sometimes modeled as \emph{mixtures} ---
weighted sums of several gaussian distributions.

As noted in \cite[Ch.11]{Murphy}, 
it is possible to shift the `clustered' (`zoned') approach in this paper
to one based on mixtures:
\begin{itemize}
 \item the EM algorithm is the standard parameter estimation method for gaussian mixtures.
 \item the $k$-means algorithm is a special case of the EM algorithm for gaussian mixtures (`hard EM')
that uses an `argmax' distance measure instead of probabilistic combination.
 \item clustered regression is a special case of what is called a \emph{mixture of experts},
in which linear regression models are combined with \emph{gating functions}
assigning weights to parts of the input space.
The resulting mixture of regression models is called a \emph{mixture regression}
\cite[Section 11.2.4]{Murphy}.
\end{itemize}

\medskip

In other words, the microclimate methods for {clusters}
can generalize directly to {mixtures}.
Although we have just presented the model formula
$~\logHEC \, \sim \, \logPHI \, \mid \, Z~$
as defining a set of separate models on zones $Z$,
mixture regression yields a single model
in which every block group has a weighted contribution to each zone. 

\medskip

A benefit of mixtures is that they can extend clustered models to larger scale.
In this perspective, microclimates are latent variables, and
the $k$-means algorithm is a \emph{vector quantization} method for finding the best single value
for these variables.  Instead, however, more general EM algorithms can obtain mixtures of regression experts  
with hidden variables \cite[Section 11.4.3--4]{Murphy}.
These algorithms could even be applied naturally at the consumer level, on the full EC dataset.

\medskip

In principle we could also generalize zones to mixtures:
each block group could have weighted participation in multiple zones.
This might give another way to move from analysis of block group aggregates
to large-scale consumer-level analysis.

However, as mentioned earlier, urban policy is based on community boundaries, not mixtures.
Urban boundaries are hard.
For example, `Piecewise Gaussian Processes' are used for urban data analysis
in \cite{PiecewiseGaussianProcesses}.
These partition the block groups into zones like our zoned Gaussian Process models.

\section{Conclusions}

In this paper we have studied annual patterns of residential electricity use in Los Angeles.
Specifically, we studied patterns as a sequence of 72 monthly values (over the six years 2006-2011)
of household electricity consumption (HEC)
in about 5000 block groups,
and obtained interesting findings about
the central importance of Household Income (PHI) on HEC,
as summarized in Section~\ref{Results}.

We found that temporal HEC patterns contain significant information about climate. 
Clustering normalized $\logHEC$ values yields clusters that correspond
both to climate and to geographic regions.
We therefore have called these regions \emph{EC-microclimate zones}.

\medskip

We showed also that EC-microclimate zones behave like environmental microclimate zones
in three perspectives:
random fields, topography, and results of $k$-means clustering.
Both kinds of zones follow a Gaussian Process Random Field (GPRF) ---
a 2D random process in which pairs of points follow Gaussian distributions,
but with covariance constrained by their geographic distance.
Microclimate variables
(wind speed, local temperature, wind pressure, and total solar irradiation)
have been modeled as Gaussian Process Random Fields in the past
\cite{UncertaintyOfMicroclimateVariables}.
However, we are unaware of other work showing that $\logHEC$ follows a GPRF.

Generally, EC-microclimate zones we have produced conform to regions
bounded by hillsides --- essentially basins, each with a distinct climate.
Furthermore, both kinds of zones also exhibit a hierarchical structure,
in which microclimates subdivide (into `submicroclimates').
Los Angeles topography and climate appears important in these subdivisions.
We have used $k$-means here because it is a common choice and
it obtains spherical clusters that appear to fit well with basin structures.
The PCA results earlier showed why normalizing an EC-pattern to total to 1 would be successful,
since patterns in a cluster have limited variation ---
so use of normalization and Euclidean clustering is justifiable.
However alternative clustering methods merit investigation.

\medskip

The rest of this paper has shown how
clustering can be a foundation for building complex models.
Clustered Regression is the natural extension of ordinary linear regression to
cope with different zones.
Clustered Gaussian Process Regression involves the same extension.
Although the data can be roughly approximated by a single Gaussian distribution,
dividing the data first by clustering into more homogeneous zones
permits better approximations and more accurate models.
This was shown in Figures~\ref{Densities} and \ref{ClusterRegression}.

We discussed at the start of this paper how prediction models are
currently of great interest in energy management.
Gaussian Process Regression models also include prediction methods that come with error bounds.
Kriging --- basically a very flexible form of interpolation --- is used heavily in geostatistics.
It can be used as a prediction method for EC, as shown in Figure~\ref{ClusterGPRegression}.
In \cite{Micromodels} we explore issues in modeling.

\medskip

There are many other directions for further work.
For example, other measures of electricity consumption similarity and variation are possible, and
other methods can be used to analyze consumption patterns.
It appears that the idea of a `EC-microclimate zone' can be useful in developing better energy policy.
In \cite{Micropolicy} we show that these zones permit useful forms of `micropolicy' (microclimate-specific policy).

\medskip

One interesting way to check the effectiveness of these zones would be to consider other urban resources --- like water.
The link between HEC and Income is consistent with published results about water consumption in Los Angeles:
affluent neighborhoods consume three times more water than others, 
and water use depends heavily on Per Capita Income \cite{WaterConsumptionPolicy}.
The three regions in Figure~\ref{basic_electricity_consumption_map} resemble the
three geographical clusters based on {water use} in \cite{WaterConsumptionPolicy}.
Considering resources like water would be a natural extension of this work.

\bigskip

\bibliographystyle{IEEEtran}
\bibliography{annoying}

\begin{IEEEbiography}[{\includegraphics[width=1in,height=1.25in,clip,keepaspectratio]{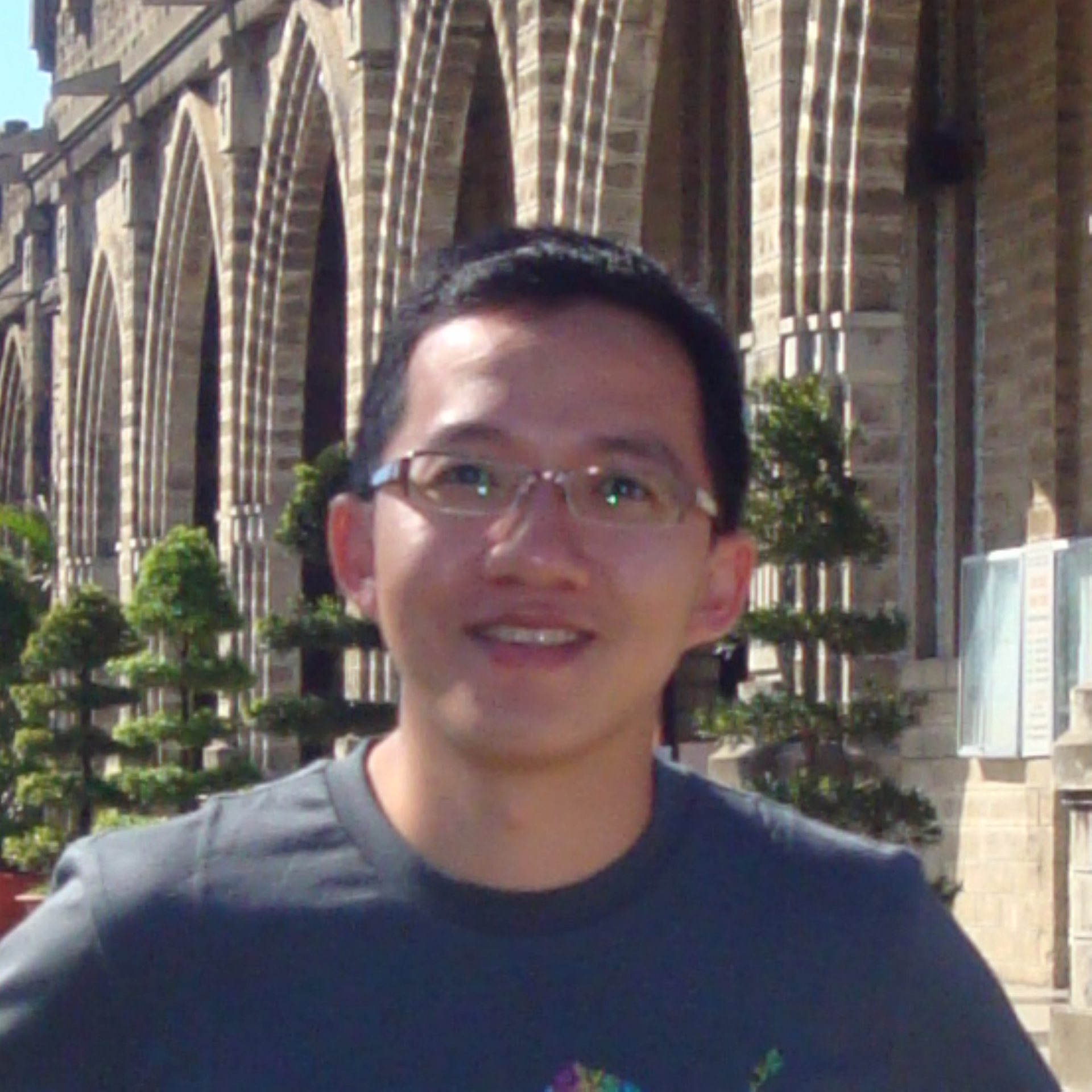}}]{Thuy Vu}
Thuy Vu is currently a graduate student in the Computer Science Department, University of California, Los Angeles.
His current research focuses on scalable learning methods for data-intensive applications in natural language processing and data mining.
He received the B.S. in Computer Science from University of Science, Vietnam in 2005,
and was a researcher at the Institute for Infocomm Research, Singapore from 2006-2011.
\end{IEEEbiography}

\begin{IEEEbiography}[{\includegraphics[width=1in,height=1.25in,clip,keepaspectratio]{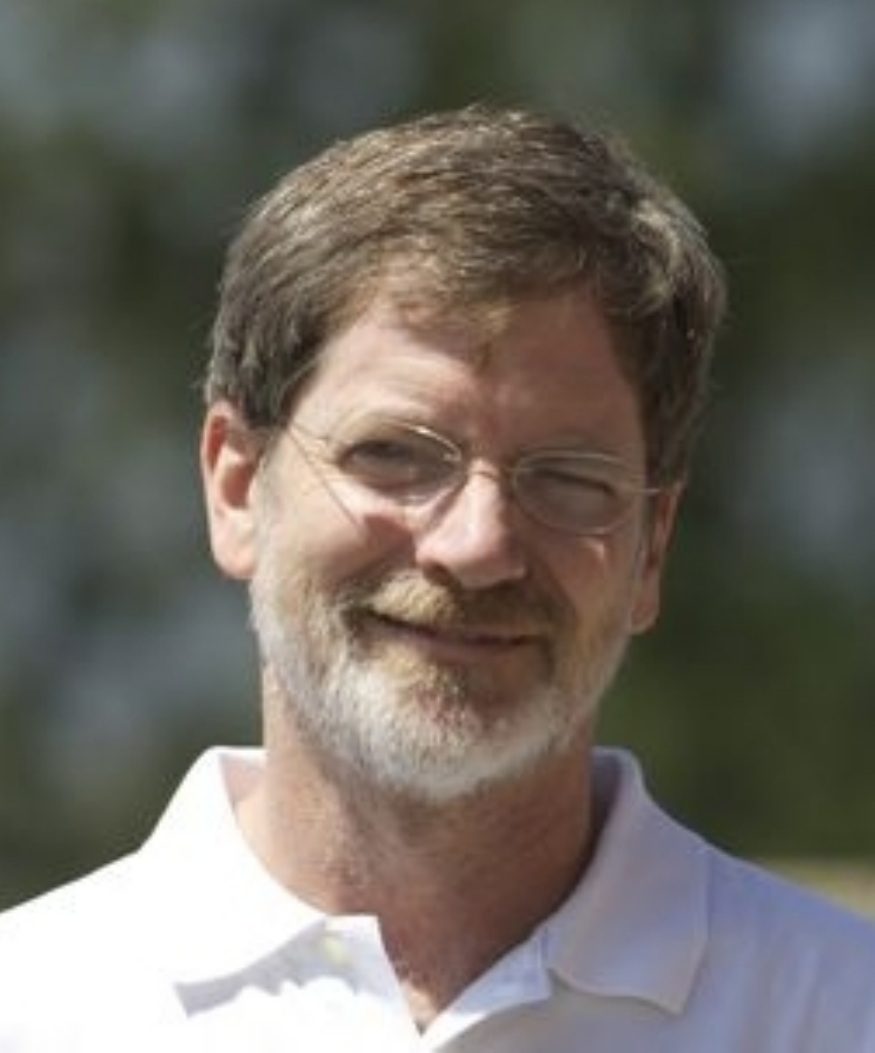}}]{D. Stott Parker}
D. Stott Parker has been on the faculty of the UCLA Computer Science Department since 1979.
He received an A.B. in Mathematics from Princeton in 1974. and a Ph.D. in Computer Science from the University of Illinois in 1978.
His interests center around data mining in general, and scientific data management in particular.
His research spans database management, knowledge base management, biomedical databases, neuroimaging, and literature mining. 
\end{IEEEbiography}

\vfill

\end{document}